%%%%%%%%%%%%%%%%%%%%%%%%%%%%%%%%%%%%%%%%%%%%%%%%%%%%%%%%%%%%%%%%%%%%%%%%%%%%
%                                                                          %
%                                                                          %
%    Macros constructed by Sourendu Gupta.                                 %
%                                                                          %
%                                                                          %
%%%%%%%%%%%%%%%%%%%%%%%%%%%%%%%%%%%%%%%%%%%%%%%%%%%%%%%%%%%%%%%%%%%%%%%%%%%%
%PAGE SIZE, FONTS, PREPRINT NUMBER ETC.
\magnification=1200\overfullrule=0pt\baselineskip=14.5pt
\vsize=23truecm \hsize=15.5truecm \overfullrule=0pt\pageno=0
\font\smallfoot=cmr10 scaled \magstep0
\font\titlefont=cmbx10 scaled \magstep1
\font\sectnfont=cmbx8  scaled \magstep1
\def\mname{\ifcase\month\or January \or February \or March \or April
           \or May \or June \or July \or August \or September
           \or October \or November \or December \fi}
\def\date{\hbox{\strut\mname \number\year}}
\def\banner{\hfill\hbox{\vbox{\offinterlineskip\crnum}}\relax}
\def\manner{\hbox{\vbox{\offinterlineskip\crnum\date}}
               \hfill\relax}
\footline={\ifnum\pageno=0\manner\else\hfil\number\pageno\hfil\fi}
%
%
% Numbering of figures, tables and equations is automatic.
%
\newcount\FIGURENUMBER\FIGURENUMBER=0
\def\FIG#1{\expandafter\ifx\csname FG#1\endcsname\relax
               \global\advance\FIGURENUMBER by 1
               \expandafter\xdef\csname FG#1\endcsname
                              {\the\FIGURENUMBER}\fi}
\def\figtag#1{\expandafter\ifx\csname FG#1\endcsname\relax
               \global\advance\FIGURENUMBER by 1
               \expandafter\xdef\csname FG#1\endcsname
                              {\the\FIGURENUMBER}\fi
              \csname FG#1\endcsname\relax}
\def\fig#1{\expandafter\ifx\csname FG#1\endcsname\relax
               \global\advance\FIGURENUMBER by 1
               \expandafter\xdef\csname FG#1\endcsname
                      {\the\FIGURENUMBER}\fi
           Fig.~\csname FG#1\endcsname\relax}
\def\figand#1#2{\expandafter\ifx\csname FG#1\endcsname\relax
               \global\advance\FIGURENUMBER by 1
               \expandafter\xdef\csname FG#1\endcsname
                      {\the\FIGURENUMBER}\fi
           \expandafter\ifx\csname FG#2\endcsname\relax
               \global\advance\FIGURENUMBER by 1
               \expandafter\xdef\csname FG#2\endcsname
                      {\the\FIGURENUMBER}\fi
           figures \csname FG#1\endcsname\ and
                   \csname FG#2\endcsname\relax}
\def\figto#1#2{\expandafter\ifx\csname FG#1\endcsname\relax
               \global\advance\FIGURENUMBER by 1
               \expandafter\xdef\csname FG#1\endcsname
                      {\the\FIGURENUMBER}\fi
           \expandafter\ifx\csname FG#2\endcsname\relax
               \global\advance\FIGURENUMBER by 1
               \expandafter\xdef\csname FG#2\endcsname
                      {\the\FIGURENUMBER}\fi
           figures \csname FG#1\endcsname--\csname FG#2\endcsname\relax}
\newcount\TABLENUMBER\TABLENUMBER=0
\def\TABLE#1{\expandafter\ifx\csname TB#1\endcsname\relax
               \global\advance\TABLENUMBER by 1
               \expandafter\xdef\csname TB#1\endcsname
                          {\the\TABLENUMBER}\fi}
\def\tabletag#1{\expandafter\ifx\csname TB#1\endcsname\relax
               \global\advance\TABLENUMBER by 1
               \expandafter\xdef\csname TB#1\endcsname
                          {\the\TABLENUMBER}\fi
             \csname TB#1\endcsname\relax}
\def\table#1{\expandafter\ifx\csname TB#1\endcsname\relax
               \global\advance\TABLENUMBER by 1
               \expandafter\xdef\csname TB#1\endcsname{\the\TABLENUMBER}\fi
             Table \csname TB#1\endcsname\relax}
\def\tableand#1#2{\expandafter\ifx\csname TB#1\endcsname\relax
               \global\advance\TABLENUMBER by 1
               \expandafter\xdef\csname TB#1\endcsname{\the\TABLENUMBER}\fi
             \expandafter\ifx\csname TB#2\endcsname\relax
               \global\advance\TABLENUMBER by 1
               \expandafter\xdef\csname TB#2\endcsname{\the\TABLENUMBER}\fi
             Tables \csname TB#1\endcsname{} and
                    \csname TB#2\endcsname\relax}
\def\tableto#1#2{\expandafter\ifx\csname TB#1\endcsname\relax
               \global\advance\TABLENUMBER by 1
               \expandafter\xdef\csname TB#1\endcsname{\the\TABLENUMBER}\fi
             \expandafter\ifx\csname TB#2\endcsname\relax
               \global\advance\TABLENUMBER by 1
               \expandafter\xdef\csname TB#2\endcsname{\the\TABLENUMBER}\fi
            Tables \csname TB#1\endcsname--\csname TB#2\endcsname\relax}
\newcount\REFERENCENUMBER\REFERENCENUMBER=0
\def\REF#1{\expandafter\ifx\csname RF#1\endcsname\relax
               \global\advance\REFERENCENUMBER by 1
               \expandafter\xdef\csname RF#1\endcsname
                         {\the\REFERENCENUMBER}\fi}
\def\reftag#1{\expandafter\ifx\csname RF#1\endcsname\relax
               \global\advance\REFERENCENUMBER by 1
               \expandafter\xdef\csname RF#1\endcsname
                      {\the\REFERENCENUMBER}\fi
             \csname RF#1\endcsname\relax}
\def\ref#1{\expandafter\ifx\csname RF#1\endcsname\relax
               \global\advance\REFERENCENUMBER by 1
               \expandafter\xdef\csname RF#1\endcsname
                      {\the\REFERENCENUMBER}\fi
             [\csname RF#1\endcsname]\relax}
\def\refto#1#2{\expandafter\ifx\csname RF#1\endcsname\relax
               \global\advance\REFERENCENUMBER by 1
               \expandafter\xdef\csname RF#1\endcsname
                      {\the\REFERENCENUMBER}\fi
           \expandafter\ifx\csname RF#2\endcsname\relax
               \global\advance\REFERENCENUMBER by 1
               \expandafter\xdef\csname RF#2\endcsname
                      {\the\REFERENCENUMBER}\fi
             [\csname RF#1\endcsname--\csname RF#2\endcsname]\relax}
\def\refand#1#2{\expandafter\ifx\csname RF#1\endcsname\relax
               \global\advance\REFERENCENUMBER by 1
               \expandafter\xdef\csname RF#1\endcsname
                      {\the\REFERENCENUMBER}\fi
           \expandafter\ifx\csname RF#2\endcsname\relax
               \global\advance\REFERENCENUMBER by 1
               \expandafter\xdef\csname RF#2\endcsname
                      {\the\REFERENCENUMBER}\fi
            [\csname RF#1\endcsname,\csname RF#2\endcsname]\relax}
\def\refs#1#2{\expandafter\ifx\csname RF#1\endcsname\relax
               \global\advance\REFERENCENUMBER by 1
               \expandafter\xdef\csname RF#1\endcsname
                      {\the\REFERENCENUMBER}\fi
           \expandafter\ifx\csname RF#2\endcsname\relax
               \global\advance\REFERENCENUMBER by 1
               \expandafter\xdef\csname RF#2\endcsname
                      {\the\REFERENCENUMBER}\fi
            [\csname RF#1\endcsname,\csname RF#2\endcsname]\relax}
\def\Ref#1{\expandafter\ifx\csname RF#1\endcsname\relax
               \global\advance\REFERENCENUMBER by 1
               \expandafter\xdef\csname RF#1\endcsname
                      {\the\REFERENCENUMBER}\fi
             Ref.~\csname RF#1\endcsname\relax}
\def\Refs#1#2{\expandafter\ifx\csname RF#1\endcsname\relax
               \global\advance\REFERENCENUMBER by 1
               \expandafter\xdef\csname RF#1\endcsname
                      {\the\REFERENCENUMBER}\fi
           \expandafter\ifx\csname RF#2\endcsname\relax
               \global\advance\REFERENCENUMBER by 1
               \expandafter\xdef\csname RF#2\endcsname
                      {\the\REFERENCENUMBER}\fi
        Refs.~\csname RF#1\endcsname{},\csname RF#2\endcsname\relax}
\def\Refto#1#2{\expandafter\ifx\csname RF#1\endcsname\relax
               \global\advance\REFERENCENUMBER by 1
               \expandafter\xdef\csname RF#1\endcsname
                      {\the\REFERENCENUMBER}\fi
           \expandafter\ifx\csname RF#2\endcsname\relax
               \global\advance\REFERENCENUMBER by 1
               \expandafter\xdef\csname RF#2\endcsname
                      {\the\REFERENCENUMBER}\fi
            Refs.~\csname RF#1\endcsname--\csname RF#2\endcsname]\relax}
\def\Refand#1#2{\expandafter\ifx\csname RF#1\endcsname\relax
               \global\advance\REFERENCENUMBER by 1
               \expandafter\xdef\csname RF#1\endcsname
                      {\the\REFERENCENUMBER}\fi
           \expandafter\ifx\csname RF#2\endcsname\relax
               \global\advance\REFERENCENUMBER by 1
               \expandafter\xdef\csname RF#2\endcsname
                      {\the\REFERENCENUMBER}\fi
        Refs.~\csname RF#1\endcsname{} and \csname RF#2\endcsname\relax}
\newcount\EQUATIONNUMBER\EQUATIONNUMBER=0
\def\EQ#1{\expandafter\ifx\csname EQ#1\endcsname\relax
               \global\advance\EQUATIONNUMBER by 1
               \expandafter\xdef\csname EQ#1\endcsname
                          {\the\EQUATIONNUMBER}\fi}
\def\eqtag#1{\expandafter\ifx\csname EQ#1\endcsname\relax
               \global\advance\EQUATIONNUMBER by 1
               \expandafter\xdef\csname EQ#1\endcsname
                      {\the\EQUATIONNUMBER}\fi
            \csname EQ#1\endcsname\relax}
\def\EQNO#1{\expandafter\ifx\csname EQ#1\endcsname\relax
               \global\advance\EQUATIONNUMBER by 1
               \expandafter\xdef\csname EQ#1\endcsname
                      {\the\EQUATIONNUMBER}\fi
            \eqno(\csname EQ#1\endcsname)\relax}
\def\EQNM#1{\expandafter\ifx\csname EQ#1\endcsname\relax
               \global\advance\EQUATIONNUMBER by 1
               \expandafter\xdef\csname EQ#1\endcsname
                      {\the\EQUATIONNUMBER}\fi
            (\csname EQ#1\endcsname)\relax}
\def\eq#1{\expandafter\ifx\csname EQ#1\endcsname\relax
               \global\advance\EQUATIONNUMBER by 1
               \expandafter\xdef\csname EQ#1\endcsname
                      {\the\EQUATIONNUMBER}\fi
          Eq.~(\csname EQ#1\endcsname)\relax}
\def\eqand#1#2{\expandafter\ifx\csname EQ#1\endcsname\relax
               \global\advance\EQUATIONNUMBER by 1
               \expandafter\xdef\csname EQ#1\endcsname
                        {\the\EQUATIONNUMBER}\fi
          \expandafter\ifx\csname EQ#2\endcsname\relax
               \global\advance\EQUATIONNUMBER by 1
               \expandafter\xdef\csname EQ#2\endcsname
                      {\the\EQUATIONNUMBER}\fi
         Eqs.~(\csname EQ#1\endcsname) and (\csname EQ#2\endcsname)\relax}
\def\eqto#1#2{\expandafter\ifx\csname EQ#1\endcsname\relax
               \global\advance\EQUATIONNUMBER by 1
               \expandafter\xdef\csname EQ#1\endcsname
                      {\the\EQUATIONNUMBER}\fi
          \expandafter\ifx\csname EQ#2\endcsname\relax
               \global\advance\EQUATIONNUMBER by 1
               \expandafter\xdef\csname EQ#2\endcsname
                      {\the\EQUATIONNUMBER}\fi
          Eqs.~\csname EQ#1\endcsname--\csname EQ#2\endcsname\relax}
\def\APEQNO#1{\expandafter\ifx\csname EQ#1\endcsname\relax
               \global\advance\EQUATIONNUMBER by 1
               \expandafter\xdef\csname EQ#1\endcsname
                      {\the\EQUATIONNUMBER}\fi
            \eqno(\APPENDIXNUMBER.\csname EQ#1\endcsname)\relax}
\def\APEQNM#1{\expandafter\ifx\csname EQ#1\endcsname\relax
               \global\advance\EQUATIONNUMBER by 1
               \expandafter\xdef\csname EQ#1\endcsname
                      {\the\EQUATIONNUMBER}\fi
            (\APPENDIXNUMBER.\csname EQ#1\endcsname)\relax}
\def\apeq#1{\expandafter\ifx\csname EQ#1\endcsname\relax
               \global\advance\EQUATIONNUMBER by 1
               \expandafter\xdef\csname EQ#1\endcsname
                      {\the\EQUATIONNUMBER}\fi
          Eq.~(\APPENDIXNUMBER.\csname EQ#1\endcsname)\relax}
%
%THESE MACROS DEFINE SECTION AND SUBSECTION HEADERS
\newcount\SECTIONNUMBER\SECTIONNUMBER=0
\newcount\SUBSECTIONNUMBER\SUBSECTIONNUMBER=0
\def\appendix#1#2{\global\let\APPENDIXNUMBER=#1\bigskip\goodbreak
     \line{{\sectnfont Appendix \APPENDIXNUMBER.\ #2}\hfil}\smallskip}
\def\section#1{\global\advance\SECTIONNUMBER by 1\SUBSECTIONNUMBER=0
      \bigskip\goodbreak\line{{\sectnfont \the\SECTIONNUMBER.\ #1}\hfil}
      \smallskip}
\def\subsection#1{\global\advance\SUBSECTIONNUMBER by 1
      \bigskip\goodbreak\line{{\sectnfont
         \the\SECTIONNUMBER.\the\SUBSECTIONNUMBER.\ #1}\hfil}
      \smallskip}
%
%DEFINE JOURNAL NAMES

\def\JSP{{\sl J.\ Stat.\ Phys.\ }}
\def\NP{{\sl Nucl.\ Phys.\ }}

\def\PR{{\sl Phys.\ Rev.\ }}

\input ../../macros/macnum
\def\crnum{\hbox{CERN-TH.6591/92 \strut}}
\def\banner{\hfill\hbox{\vbox{\crnum}}\relax}
\def\manner{\hbox{\vbox{\offinterlineskip\bigskip\bigskip
                        \crnum\date}}\hfill\relax}
\footline={\ifnum\pageno=0\manner\else\hfil\number\pageno\hfil\fi}
\def\fft{{\smallfoot Talk delivered at the Workshop on the Dynamics of
            First Order Transitions, J\"ulich, Germany, June 1--3, 1992.}}
\def\lb{\hfil\penalty-10000}
\newdimen\digitwidth\setbox0=\hbox{\rm0}\digitwidth=\wd0
\def\CHIP{\bar\chi}\def\chip{\ifmmode\CHIP\else$\CHIP$\fi}
\def\L{{\scriptscriptstyle L}}
\def\CHIP{\bar\chi}\def\chip{\ifmmode\CHIP\else$\CHIP$\fi}
\def\L{{\scriptscriptstyle L}}
{\vsize=18truecm\banner\bigskip\baselineskip=15pt
\begingroup\titlefont\obeylines
\hfil FINITE-SIZE SCALING\hfil
\hfil ON THE ISING COEXISTENCE LINE\hfil
\endgroup\bigskip
\bigskip\centerline{Sourendu Gupta\footnote{${}^*$}{\fft}
and A.~Irb\"ack}\bigskip
\centerline{Theory Division, CERN, CH-1211, Geneva 23, Switzerland.}
\bigskip\bigskip\bigskip\centerline{\bf ABSTRACT}\medskip
We report tests of finite-size scaling ansatzes in the low temperature phase
of the two-dimensional Ising model. For moments of the magnetisation density,
we find good agreement with the new ansatz of Borgs and Koteck\'y, and clear
evi
consequences of the convexity of the free energy are not adequately treated
in either of these approaches.\lb
\vfil\eject}
We report some simulations of the 2-d Ising model on the phase coexistence
line, $\beta>\beta_c$ and $h=0$. The motivation for this work arises from some
recent developments in the theory of finite-size scaling (FSS) at phase
coexistence. Early work centred around a phenomenological model,$^1$
the double Gaussian model (2G). This emphasised the role of the peaks of
probability distributions (pure phases) but completely neglected interfacial
phenomena due to coexistence. A different ansatz (BK), based on a
rigorous computation for the 2-d Ising model at large $\beta$ and 2-d $q$-state
Potts models at large $q$, has now been developed.$^2$
This proceeds by identifying contributions to the
partition function due to small fluctuations around pure phases and bounding
the remainder. It provides predictions for the shift in the transition
coupling and finite-size effects in moments of the energy or order-parameter.
BK can be used to approximately justify 2G.
Results of very large scale computations with 2-d Potts models ($q>4$) have
not been totally consistent with these predictions.$^3$

The Ising coexistence line is an obvious test-bed for high-precision numerical
work on this problem.
With any definition of the pseudocritical point which respects
the $Z_2$ symmetry of the system,
at all volumes, the position of this line is given
precisely by the condition $h=0$. The correlation length can be tuned easily
by changing $\beta$. The symmetry relation between phases reduces the
number of free parameters. The cluster algorithm allows easy simulation of
the theory, with efficient sampling of the different phases, as well as the
small fluctuations around them which are responsible for the finite-size
corrections. Our results are consistent with the predictions of BK,
and accurately reflect the drawbacks of 2G. The small but definite
discrepancies are similiar to those seen in Potts simulations. These are due
to the convexity of the free energy in the thermodynamic limit.

The critical point of the Ising model is at $\beta_c=\ln(\sqrt2+1)\approx
0.44$. For $\beta>\beta_c$, the spontaneous magnetisation density $m_0$
is given by
$$m_0\;=\;\left(1-{1\over\sinh^42\beta}\right)^{1/8}.  \eqno(1)$$
The correlation length is given by
$$\xi^{-1}\;=\;2\sqrt2\ln\sinh2\beta.     \eqno(2)$$
A low-temperature expansion for the pure phase susceptibility \chip{} is
known to 9 terms$^4$ in $\exp(-4\beta)$, and can be extended by methods
outlined in.$^5$ The couplings at which our simulations were
performed are listed in Table 1, along with the lattice sizes and
statistics used, and the values of $m_0$ and $\xi$ at these couplings.
In all results reported here, errors have been estimated through a jack-knife
procedure based on partitioning the data into 1000 blocks. The first $10^5$
sweeps have been discarded for thermalisation. Autocorrelations in the data
are negligible.

{\midinsert
Table 1. The couplings $\beta$ at which our simulations have been performed
alon
the exact values of $m_0$ and $\xi$. Also listed are the largest lattice size
$L_{max}$ and the number of Swendson-Wang sweeps $N_{sw}$ (in millions). The
statistics increases with the lattice size.
{$$\vbox{\offinterlineskip\halign{
      \vrule#&\strut\hfil$\;#\;$\hfil&
      \vrule#&\hfil$\;#\;$\hfil&\vrule#&\hfil$\;#\;$\hfil&
      \vrule#&\hfil$\;#\;$\hfil&\vrule#&\hfil$\;#\;$\hfil&
      \vrule#&\hfil$\;#\;$\hfil&\vrule#\cr
\noalign{\hrule}
&2\beta&&m_0&&\xi&&L_{max}&&N_{sw}&\cr
       \noalign{\hrule}
&0.95&&0.86578&&3.73&&128&&3.6{\rm--}10&\cr
&1.00&&0.91132&&2.19&&128&&2.0{\rm--}5&\cr
&1.10&&0.95394&&1.22&&64&&2.0{\rm--}4&\cr
       \noalign{\hrule}}}$$}
\endinsert}

The probability distribution in the magnetisation density $P_\L(m)$ can be used
to extract a free energy
$$F_\L(m)\;=\; -{1\over L^2}\ln P_\L(m) + c_\L,      \eqno(3)$$
where the constant is chosen such that the minimum value of $F_\L(m)$ is zero
for all lattice sizes $L$. In the thermodynamic limit $L\rightarrow\infty$,
$F(m)$ has been proved to be convex.$^6$ The development of the
convexity is
illustrated in Fig.~1. The symmetry of the problem has been used to
show only the branch for $m>0$. There is evidence of a shift in the position of
the minimum. However, the granularity in $m$ prevents us from making an
accurate measurement of the magnitude of the shift. The analogous shift in
the latent heat for Potts models$^7$ has been attributed to the
developing convexity of the free energy.

We have extracted expectation values for the moments of $m$. Odd moments are
zero due to symmetry. Even moments are therefore equivalent to central moments.
Finite-size corrections to these can be organised in the series
$$\langle m^{2k}\rangle\;=\;
     m_0^{2k} + {a_k\over L^2} + {b_k\over L^4} + \cdots,  \eqno(4)$$
where exponentially small terms have been neglected.
In both 2G and BK, the coefficient of the leading correction is given by
$$a_k\;=\; (2k-1)k m_0^{2k-2} {\chip\over\beta},   \eqno(5)$$
and $b_1=0$.

Due to terms neglected in the quadratic approximation,
2G cannot be expected to give correct results for
$b_k$ or higher order coefficients in the series in (4). If one
nevertheless applies this probability distribution, then
one gets coefficients, given in terms of
$k$, $m_0$ and \chip, which are all non-negative.

We tried to fit the first two terms in the series to our data.
The results for the
first four even moments are shown in Fig.~2.
The error bars decrease with increasing order $k$, which we
attribute simply to the normalisation of the operator studied.
At large $L$, we find that the data are well described by the
two first terms. Data points at smaller $L$ lie
below the fitted lines, which shows
that 2G, indeed, does not produce the correct higher order corrections.
This behaviour could have been anticipated directly from
the probability distribution in Fig.~1,
which shows that $F(m)$ is not quadratic
close to its minimum.
The shallower curvature on the small $m$
side pulls the moments below the 2G prediction.
To obtain more detailed information about the shape
of the distribution and the terms neglected in the quadratic
approximation, we have also investigated
single phase cumulants.$^8$

BK and 2G have, as mentioned above, the same
$1/L^2$ term,
whereas higher order corrections differ in general.
In BK the latter are not given in terms of $m_0$ and \chip\
only, but new parameters
enter. The coefficient $b_2$, for example,
depends in addition on the single phase cumulant of third order.
Our result for this cumulant gives a negative sign and
a reasonable magnitude for $b_2$. The finite-size behaviour of the
measured moments are completely consistent with BK.
However, in the limit of large
$L$, the region of $m$ between
the minima of $F(m)$ must scale to zero. This feature of the free energy is
not described by BK.

{\midinsert
Table 2. The spontaneous magnetisation extracted at each $\beta$ from the
expect
value of the $2k$-th moment of the magnetisation.
{$$\vbox{\offinterlineskip\halign{
      \vrule#&\strut\hfil$\;#\;$\hfil&
      \vrule#&\hfil$\;#\;$\hfil&\vrule#&\hfil$\;#\;$\hfil&
      \vrule#&\hfil$\;#\;$\hfil&\vrule#&\hfil$\;#\;$\hfil&
      \vrule#&\hfil$\;#\;$\hfil&\vrule#\cr
\noalign{\hrule}
&2\beta&&k=1&&k=2&&k=3&&k=4&\cr
       \noalign{\hrule}
&0.95&&0.86581(2)&&0.86582(2)&&0.86582(2)&&0.86584(2)&\cr
&1.00&&0.91135(2)&&0.91135(2)&&0.91136(2)&&0.91137(2)&\cr
&1.10&&0.95399(3)&&0.95399(3)&&0.95399(3)&&0.95399(3)&\cr
       \noalign{\hrule}}}$$}
\endinsert}

The 2-parameter fit to the variation of moments yields values of $m_0$ in
excellent agreement with the exact results. These are shown in Table 2.
The relation of (5) implies a relation between the coefficients of
the fits to different moments. As shown in Table 3, these are true to
reasonable accuracy. The agreement between the coefficients becomes better
if the moments for $k\ge2$ are fitted to three terms of the series in
(4). This agreement encourages us to extract a pure-phase susceptibility
from the fits. These are shown in Fig.~3 along with the results obtained
from a series expansion (as described earlier). The agreement is rather
striking.

{\midinsert
Table 3. Check of the relation in (5) for the data taken at $2\beta=0.95$. The
value of $a_k$ predicted from $a_l$ and
the $m_0$ obtained from the $l$-th
moment $(l<k)$. The diagonal entries are the direct result of the fit. The
error estimates for the off-diagonal entries take into account the covariance
of the fitted parameters.
{$$\vbox{\offinterlineskip\halign{
      \vrule#&\strut\hfil$\;#\;$\hfil&
      \vrule#&\hfil$\;#\;$\hfil&\vrule#&\hfil$\;#\;$\hfil&
      \vrule#&\hfil$\;#\;$\hfil&\vrule#&\hfil$\;#\;$\hfil&
      \vrule#&\hfil$\;#\;$\hfil&\vrule#\cr
\noalign{\hrule}
&l&&k=1&&k=2&&k=3&&k=4&\cr
       \noalign{\hrule}
&1&&2.3(2)&& && && &\cr
&2&&10.3(9)&&9.6(3)&& && &\cr
&3&&19.(2)&&18.0(5)&&17.0(3)&& &\cr
&4&&27.(2)&&25.1(7)&&24.3(5)&&23.8(4)&\cr
       \noalign{\hrule}}}$$}
\endinsert}
\vfil\eject
\bigskip\centerline{\sectnfont REFERENCES}\bigskip
\item{1)}
    K.~Binder and D.~P.~Landau, \PR B30 (1984) 1477;\lb
    M.~S.~S.~Challa, D.~P.~Landau and K.~Binder, \PR B34 (1986) 1841;\lb
    See also K.~Binder, these proceedings.
\item{2)}
    C.~Borgs and R.~Koteck\'y, \JSP 61 (1990) 79;\lb
    C.~Borgs, R.~Koteck\'y and S.~Miracle-Sol\'e, \JSP 62 (1991) 529;\lb
    See also C.~Borgs, these proceedings.
\item{3)}
    A.~Billoire, R.~Lacaze and A.~Morel, \NP B370 (1992) 773;\lb
    See also A.~Billoire, these proceedings.
\item{4)}
    C.~Domb, in {\sl Phase Transitions and Critical Phenomena}, Vol 3,
    Eds.\ C.~Domb and M.~S.~Green, Academic Press, London, 1974.
\item{5)}
    D.~S.~Gaunt and A.~J.~Guttmann, in {\sl Phase Transitions and Critical
    Phenomena}, Vol 3, Eds.\ C.~Domb and M.~S.~Green, Academic Press,
    London, 1974.
\item{6)}
    R.~B.~Griffiths, \PR 152 (1966) 240.
\item{7)}
    J.~Lee and J.~M.~Kosterlitz, \PR B43 (1991) 3265.
\item{8)}
    S.~Gupta, A.~Irb\"ack and M.~Ohlsson, in preparation.
\vfil\eject
\bigskip\centerline{\sectnfont FIGURE CAPTIONS}\bigskip
\item{Figure 1:}
   The free energy $F(m)$ extracted from the distribution of
   magnetisation density at $2\beta=0.95$ on $128^2$ (circles), $96^2$
   (open squares), $64^2$ (filled squares) and $40^2$ (triangles) lattices.
\item{Figure 2:}
   FSS of the first 4 even moments of magnetisation at $2\beta=0.95$.
   2-parameter fits (lines) have been made to data from the 5 largest
   lattices (filled circles). The downward curvature of the data on the
   smaller lattices (open circles) cannot be accounted for within
   2G and is due to the developing convexity of $F(m)$.
\item{Figure 3:}
   The pure phase susceptibility obtained from the 6-th (open circles) and
   8-th (filled circles) moments of the magnetisation density along with the
   results obtained from the first 9 terms of the low temperature series
   (full line) and the extrapolated series expansion (broken line).
\vfill\end